\documentclass[aps,prd,twocolumn,floats,nofootinbib,longbibliography]{revtex4-1}
\usepackage{graphicx}
\usepackage{graphics}
\graphicspath{{./figs/}}
\usepackage{bm}
\usepackage{amsmath,amssymb}
\usepackage{aas_macros}
\usepackage{calrsfs}
\usepackage{natbib}
\usepackage[colorlinks=true]{hyperref}
\usepackage{breqn}
\usepackage{color}
\usepackage{changes}
\usepackage[T1]{fontenc}
\usepackage{ae,aecompl}
\definechangesauthor[name={Leo}, color=orange]{LB}
\definechangesauthor[name={Agnes}, color=teal]{AF}
\definechangesauthor[name={Olivier}, color=magenta]{OM}
\definechangesauthor[name={Jacques}, color=red]{JL}

\def\aap{\textit{Astronomy and Astrophysics}}

\def\be{\begin{equation}}
\def\ee{\end{equation}}
\def\bea{\begin{eqnarray}}
\def\eea{\end{eqnarray}}

\def\lg{\lambda_g}

\usepackage{graphicx}	
\usepackage{amsmath}	
\usepackage{amssymb}	
\usepackage{changes}

\def\Lprl{$1.83 \times 10^{13}$ km} 
\def\mprl{$6.76 \times 10^{-23}$ eV/$c^2$}

\def\Lnonenteb{ $1.82 \times 10^{13}$ km} 

\def\Lnonente{ $3.93 \times 10^{13}$ km} 
\def\mnonente{$3.16 \times 10^{-23}$ eV$/c^2$}

\def\LnonenteNeg{ $3.77 \times 10^{13}$ km} 

\def\Ltrois{ $3.43 \times 10^{13}$ km} 
\def\mtrois{$3.62 \times 10^{-23}$ eV$/c^2$}

\def\LtroisNeg{ $3.02 \times 10^{13}$ km} 

\def\Lcinq{ $2.69 \times 10^{13}$ km} 
\def\mcinq{$4.61 \times 10^{-23}$ eV$/c^2$}

\def\LcinqNeg{ $2.82 \times 10^{13}$ km} 

\def\clnonente{90\% C.L}
\def\cltrois{99.7\% C.L}

\begin{document}
\title{Constraint on the Yukawa suppression of the Newtonian potential from the planetary ephemeris INPOP19a}

\author{
L. Bernus $^{1,2}$,
O. Minazzoli $^{3,4}$,
A. Fienga $^{2,1}$,
M. Gastineau $^{1}$,
J. Laskar $^{1}$,
P. Deram $^2$,
A. Di Ruscio$^{2,5}$\\
$^{1}$IMCCE, Observatoire de Paris, PSL University, CNRS, Sorbonne Universit\'e, 77 avenue Denfert-Rochereau, 75014 Paris, France \\
$^{2}$G\'eoazur, Observatoire de la C\^ote d'Azur, Universit\'e C\^ote d'Azur, IRD, 250 Rue Albert Einstein, 06560 Valbonne, France\\
$^{3}$Centre Scientifique de Monaco, 8 Quai Antoine 1er, Monaco \\
$^{4}$Artemis, Universit\'e C\^ote d'Azur, CNRS, Observatoire de la C\^ote d'Azur, BP4229, 06304, Nice Cedex 4, France\\
$^{5}$ Dipartimento di Ingegneria Meccanica e Aerospaziale, Sapienza Universit\`a di Roma, via Eudossiana 18, 00184 Rome, Italy\\
}

\begin{abstract}
We use the latest solution of the ephemeris INPOP (19a) in order to improve our previous constraint on the existence of a Yukawa suppression to the Newtonian potential, generically associated to a graviton's mass. Unlike the ephemeris INPOP17a, several residuals are found to degrade significantly at roughly the same amplitudes of the Compton wavelength $\lg$. As a consequence, we introduce a novel statistical criterion in order to derive the constraint with INPOP19a. After checking that it leads to a constraint consistent with our previous result when applied on INPOP17b, we apply the method to the new solution INPOP19a.  We show that the residuals of Mars orbiters, Cassini, Messenger, and Juno, degrade significantly when $\lambda_g \leq$ \Ltrois \ with a 99,7\% confidence level---corresponding to a graviton mass bigger than \mtrois. This is a stronger constraint on the Compton wavelength than the one obtained from the first gravitational-wave transient catalog by the LIGO-Virgo collaboration in the radiative regime, since our \clnonente. limit reads $\lambda_g >$\Lnonente~($m_g <$~\mnonente).
\end{abstract}
\maketitle

\section{Introduction}
\label{sec:intro}
It has long been suggested that solar system observations could be used to put stringent constraints on the existence of a Yukawa suppression of the Newtonian potential \cite{will:1998pr}, which is generically associated to the hypothesis that the gravitational field may possess a mass \cite{derham:2014lr,derham:2017rp}. In a recent publication \cite{bernus:2019prl}, we used the planetary ephemeris INPOP17b to constrain the Compton wavelength in the Yukawa suppression to be larger than \Lprl~at the \clnonente --- corresponding to a mass of the gravitational field smaller than \mprl. INPOP (Int\'egrateur Num\'erique Plan\'etaire de l'Observatoire de Paris) \cite{fienga:2008aa} is a planetary ephemeris that is built by integrating numerically the equations of motion of the main planets, the Moon and 14000 asteroids of our the Solar System following the formulation of \cite{moyer:2003}, and by adjusting to Solar System observations such as lunar laser ranging or space mission observations. With the constant improvement of the solar system model and the permanent addition of new data, new versions of the ephemeris are derived and distributed on a regular basis since 2006 \citep{fienga:2008aa,fienga:2011cm,fienga:2015cm,fienga:2019inpop}. 
The new solution INPOP19a benefits from the use of 9 Jupiter very accurate positions deduced from the Juno mission. It also benefits from new Cassini data including i) data obtained during the final phase of the mission in 2017 (labelled {\it{Grand Finale}} here) and ii) Navigation and Gravity flybys data obtained between 2006 and 2016 (labelled {\it{r}} in the following). Both data sets were analysed from raw recordings by our team, focusing specifically on uncertainty estimation. In INPOP17b, Cassini data were also used but with a more limited time coverage (from 2004 to 2014) and were produced by JPL \citep{hees2014prd} but without a clear estimation of the uncertainties. In \cite{DiRuscio2020}, one will find the details on our Cassini data analysis and on deduced uncertainties whereas full comparisons with INPOP17a are given in \cite{fienga:2019inpop}. INPOP19a shows also improvements relative to INPOP17b thanks to the introduction in the dynamical modeling of the perturbation of a Trans-Neptunian Objects (TNO) ring \cite{DiRuscio2020}. 
Indeed, as explained in \cite{DiRuscio2020, fienga:2020aap}, the use in INPOP of the very accurate positions of Saturn obtained during the last phase the Cassini mission (Grand Finale), requires the introduction in the dynamical modeling of the perturbations induced by Trans-Neptunian objects for a better representation of the Grand Finale observations. Such an introduction has been done in using a ring potential with a mass estimated with the INPOP19a adjustement.The estimation of this mass has been published in \cite{DiRuscio2020} and constraints on the unknown P9 planets have also been obtained with this modeling  \citep{fienga:2020aap}
Overall, by comparing with INPOP17b, the accuracy of Jupiter orbit was improved by two orders of magnitudes. For Saturn, the improvement between INPOP17a and INPOP19a is of a factor 30 for the Grand Finale and 2.6 for the period between 2006 and 2016. More details of this update can be found in\cite{DiRuscio2020, fienga:2020aap}.

Statistics on the residuals of various INPOP solutions are sometimes used in order to derive constraints on various alternative theories, e.g. \cite{will:2018cq}. However, we explained in our previous publication \cite{bernus:2019prl} that doing so tends to overestimate the constraints on alternative theories. The reason being that the fit to the data is model dependent. Hence, the fitted parameters of the INPOP ephemerides are in general valid for general relativity only. Meaning that for alternative theories, the best fit of the parameters will in most cases be different from the ones obtained while assuming general relativity. Alternative theory parameters are indeed usually highly correlated to other parameters of the ephemeris \cite{bernus:2019prl}, such that potential effects of alternative theories could be in part absorbed by the modification of other parameters of the ephemeris. Therefore, a rigorous estimation of the constraint on alternative theory parameters calls for a new global fit of the ephemeris parameters to the data in the framework of the considered alternative theory. 

To do so in the framework of a Yukawa suppression of the Newtonian potential, we start off with an ephemeris that assumes general relativity (INPOP17b or 19a). We then add the extra acceleration term that is due to the Yukawa suppression to the numerical integrator, which, at the leading perturbative order, reads \cite{bernus:2019prl} 
\begin{equation}
\label{eq:accgravmas}
	\delta a^i = \frac{1}{2} \sum_P \frac{G M_P}{\lg^2} \frac{x^i-x_P^i}{r}, 
\end{equation}
for a set of values of $\lg$ between $10^{13}$ and $10^{14}$ km . Next we re-perform a global fit of the INPOP parameters to the data according to the procedure described in \cite{viswanathan:2018dc} for each value of $\lg$. We then evaluate statistically the level of $\lg$ at which the residuals are too degraded with respect to the original solution.
In this work, we presents results obtained while using the latest updated ephemeris, INPOP19a.
Contrary to what happened with INPOP17b for which only the residuals deduced from the Cassini radio experiment were significantly degraded with the considered values of $\lg$, residuals from several different data sets and planets---not only Cassini and Saturn---degrade simultaneously with INPOP19a (see Fig. \ref{fig:residuals}). Therefore, in order to take into account all the degradations simultaneously, we derived a new statistical method with respect to our previous publication \cite{bernus:2019prl}, which we shall present now.

\section{Global observational $\chi^2$ constraint}
\label{sec:like}

Now, lets consider that the global fit of the ephemeris parameters has already been performed, such that the residuals are minimized for each value of $\lg$.
Let us consider such a solution for a given value of $\lg$ and we then define its observational $\chi^2$
\begin{equation}
	n\chi^2(\lg) = \sum_{\Omega_j} n_j \frac{\sigma_j^2(\lg)}{\sigma_{o,j}^2}
\end{equation}
where $\Omega_j$ are the different observational sets that we get from different sources (Cassini, Messenger, optical etc.), $n_j$ is the number of points of each data sets, $n=\sum_jn_j$ is the total number of observations, $\sigma_{o,j}$ is the experimental uncertainty of observation $j$, and $\sigma_j^2(\lg)$ is the standard dispersion of the residuals between the simulated observables with a Yukawa suppression with a Compton wavelength $\lg$ and the observations of the set $j$. We note $\chi^2(\infty)=\chi^2_r$, the observational $\chi^2$ of the reference solution. If the residuals are in a linear vicinity of $0$ and follow a Gaussian distribution,  $n\chi^2$ follows a $n$ degrees of freedom $\chi^2$ law, and that when $n\rightarrow\infty$ 
\begin{equation}\label{eq:thm}
	z(\lg)=\sqrt{\frac{n}{2}}(\chi^2(\lg)-\chi^2_r) \xrightarrow[n \to \infty]{} \mathcal N (0,1)
\end{equation}
In other words, $z$ tends to follow a normal distribution centered around 0 and of standard dispersion equal to 1. 

However, we have tested that some observations are almost insensitive to the Yukawa suppression, so we don't need to compute all the simulations of the observations. We show in the next section how sensitive or non-sensitive observations were chosen in this study. According to each specific alternative theory, each ephemeris can have its sensitive and non-sensitive observational data and they should all be analysed case-by-case.
In INPOP, the weights correspond to the observational uncertainties.
Under these assumptions, let $\Omega$ be the set of all the observations, and $\tilde{\Omega}$ the set of sensitive observations with $\tilde{n}$ the total number of sensitive observations, and 
\begin{equation}
	\tilde{n}\tilde{\chi}^2=\sum_{j\in\tilde{\Omega}}n_j\frac{\sigma_j^2(\lg)}{\sigma_{o,j}^2} \label{eq:chi2tilde}
\end{equation}
Let us assume that if $j$ is a non sensitive observational dataset, we have $\sigma_j(\lg)=\sigma_{r,j}$ (where the label $r$ indicates the reference ephemeris , here INPOP19a). Under these assumptions, a straightforward calculation leads to
\begin{equation}\label{eq:deltachi21}
	\chi^2(\lg)-\chi^2_r = \frac{\tilde{n}}{n}(\tilde{\chi}^2(\lg)-\tilde{\chi}^2_r)
\end{equation}
However, it appears that the observational uncertainties of the sensitive observations are very close to the standard dispersions of the reference solution, such that we can set 
\begin{equation}
	\forall j\in\tilde{\Omega}, \sigma_{r,j}=\sigma_{o,j}
\end{equation}
without loss of generality. Indeed, for the reference solution (here INPOP19a), we have $\chi^2_r\approx 1.0036$. A straightforward differential calculation shows that in order to get $\chi^2_r=1$ modifying only the observational uncertainties, each $\sigma_{o,i}$ should be modified as follows
\begin{equation}
	\delta \sigma_{o,i} = (\chi^2_r-1)\frac{n_i\sigma_{r,i}^2}{\sigma_{o,i}^3}\left(2\sum_{\Omega_i}n_i^2\sigma_{r,i}^4/\sigma_{o,i}^6\right)^{-1}.
\end{equation}
Since we have $\sigma_{r,i}\approx\sigma_{o,i}$, we get
\begin{equation}
	\delta \sigma_{o,i} \approx \frac{1}{2}(\chi^2_r-1)\left( 1+ \sum_{j\ne i}\frac{n_j\sigma_{o,i}^2}{n_i\sigma_{o,j}^2} \right)^{-1} \le \frac{\chi^2_r -1}{2} \approx 0.2\%.
\end{equation}
Therefore, modifying $\sigma_{o,i}$ by less than $0.2\%$ would be enough to set $\chi^2_r$ to 1. On another hand, we know that the observational uncertainties are difficult to estimate with accuracy as they are the combination of several contributions (instrumental uncertainties, calibrations, uncomplete modelings etc...). Therefore it is legitimate to set $\tilde{\chi}^2_r=1$ in what follows. Eq. \eqref{eq:deltachi21} then becomes
\begin{equation}
	\chi^2(\lg)-\chi^2_r = \frac{\tilde{n}}{n}(\tilde{\chi}^2(\lg)-1).
\end{equation}
From here, using \eqref{eq:thm}, we see that 
\begin{equation}
	z(\lg)=\sqrt{\frac{\tilde{n}}{2}}(\tilde{\chi}^2(\lg)-1), \label{eq:zlambda}
\end{equation}
follows a 0-centered normal distribution of dispersion equal to 1. We can then compute the likelihood of each ephemeris :
\begin{equation}
	L(\lg) = 1 - \frac{1}{\sqrt{2\pi}}\int_{-\infty}^{z(\lg)}\exp\left(-\frac{x^2}{2}\right)\ dx \label{eq:likelihood}.
\end{equation}
This quantity is interpreted as the probability of a tested theory to be likely---here, the Yukawa suppression of the Newtonian potential. For the reference solution, we have $L_r=1/2$. This means that a tested theory which reproduces the same $L$ that the reference solution has as many chances to be a better theory than a worse one, with respect to the reference solution. In this case, the residuals are the same or very close to the one of the reference. If $L>1/2$, the theory is said to be better than the reference solution, with smaller residuals. If $L<1/2$, the theory is said to be worse, following a degradation of the residuals. With this method we can detect if some values of $\lg$ improve or degrade the residuals. If $L$ becomes very close to 1, then the theory is said to be much better than the reference theory. In order to compare our result to the existing literature, we consider a \clnonente \ criterion, corresponding to $L(\lg)<0.1$. Nevertheless, as for the classical gaussian distributed variable, we also take the equivalent of the $3$-$\sigma$ criterion : theories for which $L(\lg)<0.003$ will be rejected with a probability of 99.7\%. Table \ref{tab:resultats} gives the constraints on $\lg$ according to different criteria.

\section{Results and discussion}
\subsection{Results with INPOP17b}

\begin{figure}
	\includegraphics[scale=0.6]{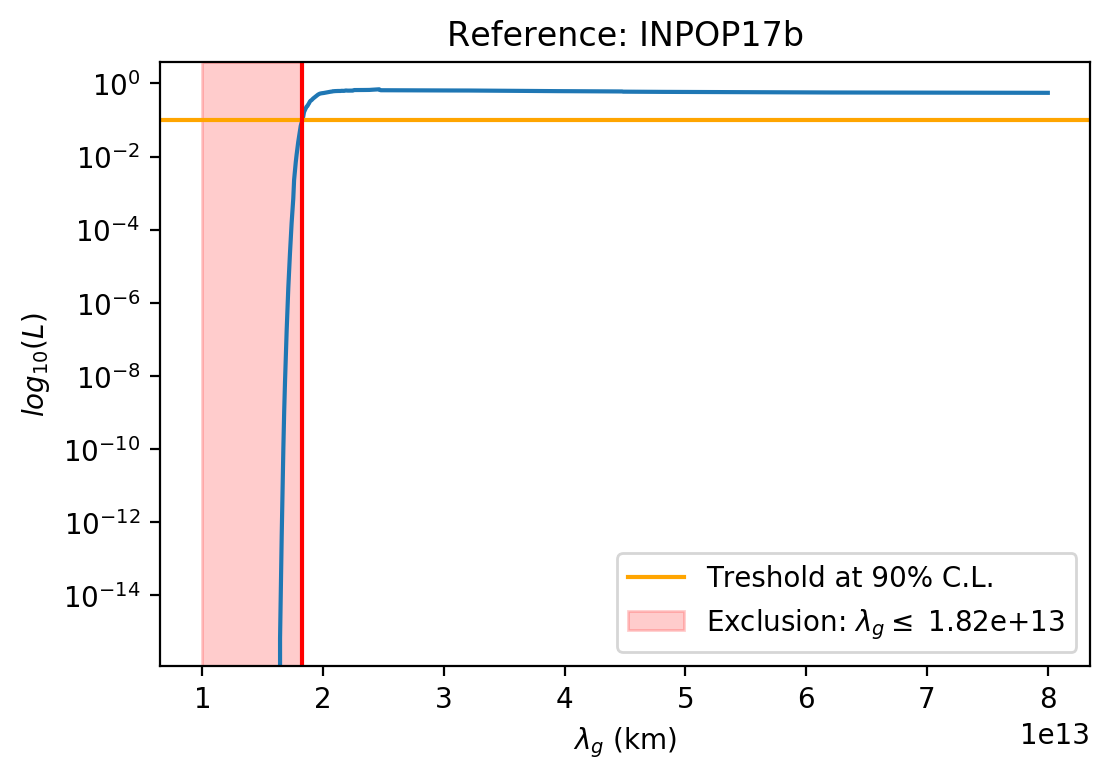}
	\caption{Likelihood for each value of $\lg$ of the Yukawa suppression deduced from INPOP17b. The horizontal segment represents the 90\% confidence limit; while the red vertical segment indicates the value of $\lg$ at that limit. The shaded area is the exclusion region. The likelihood is given for each point of the considered set of values for $\lg$. In order to be conservative, we identify first the lowest value of $\lg$ such that $L(\lg)$ is still above the threshold, and then we consider the next (lower) value to this one as our constraint. The same procedure is applied in figures \ref{fig:likelihood} and \ref{fig:gravmass_moins}. The constraint reads $\lg >$\Lnonenteb \ at the \clnonente.}
	\label{fig:likelihood17b}
\end{figure}

Before using this test with INPOP19a, let us confirm that the new criterion defined in the previous section gives consistent results compared to the one obtained with INPOP17b \cite{bernus:2019prl}. We had already computed the residuals standard dispersion with respect to $\lg$.
With INPOP17b the Cassini residuals are the by far the most sensitive to the Yukawa suppression (see Fig. 1 of the Supplemental Materials of \cite{bernus:2019prl}). When the the residuals of the other data begin to increase, the ones from Cassini are already too big to be acceptable. Therefore, the sensitive $\tilde{\chi}^2(\lg)$ is limited to Cassini data and Eq. \eqref{eq:zlambda} reduces to
\begin{equation}
	z(\lg)=\sqrt{\frac{\tilde{n}}{2}}\left(\frac{\sigma_{Cassini}^2(\lg)}{\sigma_{Cassini}^2(ref)}-1\right),
\end{equation}
where $\tilde{n}$ is the number of observations of Cassini in INPOP17b and $\sigma_{Cassini}(ref)$ is the INPOP17b standard deviation for Cassini. From the statistics of the residuals in \cite{bernus:2019prl}, one can directly deduce the likelihood of the massive graviton when INPOP17b is taken as a reference solution. We show this in Fig. \ref{fig:likelihood17b}, where we focus on the \clnonente \ limit. Our new statistical method applied on the residuals obtained in \cite{bernus:2019prl} lead to $\lambda_g > 1.82 \times 10^{13}$\,km at the \clnonente; whereas the method that was used in \cite{bernus:2019prl} led to $\lambda_g > 1.83 \times 10^{13}$\,km at the \clnonente. As the results obtained with the two criteria are consistent, we can state that the likelihood criterion described above is validated.

\subsection{Results with INPOP19a}

Following the same procedure as explained in \cite{bernus:2019prl}, we computed 576 ephemerides with fixed values of $\lg$ from $10^{13}$km to $10^{14}$km, fitting all the other parameters of the model for finding the optimal solution. For each of them we computed $\tilde{\chi}^2(\lambda)$ defined in Eq. \eqref{eq:chi2tilde} and then deduced the likelihood. We do 26th iteration for each value of $\lg$ in order to have a convergence of the parameters in each case. The sensitive observations that were retained for the computation of $L(\lg)$ are summarized in Table \ref{tab:sensdata}.  All the data used for the adjustment procedure, sensitive or not, are listed in the documentation of INPOP19a \cite{fienga:2019inpop}.  To have the same kind of data in the statistical computation, we computed a daily average of Mars Express and Mars Odyssey such that there are 5993 Marsian points. From this and Table \ref{tab:sensdata} we get $\tilde{n}=7856$ and accordingly we can compute $L(\lg)$ of Eq. \eqref{eq:likelihood}.

\begin{table}
	\caption{Summary of the the data sets and their average observational uncertainties $\sigma_{r}$, in meters. Messenger data where provided by \cite{verma2016jgr}. ``Cassini JPL'' data are those given by JPL \citep{hees2014prd}. Cassini Navigation and Gravity flybys data and Grand Finale are those reduced by our team \cite{fienga:2019inpop,DiRuscio2020}.}

	\begin{tabular}{cccc}
		Observations&  \# & dates & $\sigma_r$ (m) \\
		\hline
		Messenger & 1065&2011-2014 &4.1 \\
		Mars Express &  27849 &  2005-2017 &      2.0  \\
		Mars Odyssey &  18234 &  2002-2014 &      1.3  \\
		Cassini JPL &166 &2004-2014 & 25 \\
		Cassini Navigation and Gravity flybys & 614 & 2006-2016 & 6.1\\
		Cassini Grand Finale & 9 & 2017 & 2.7 \\
		Juno & 9 &  2016-2018 &      18.5 %
	\end{tabular}

	\label{tab:sensdata}
\end{table}

\begin{figure}
	\includegraphics[scale=0.5]{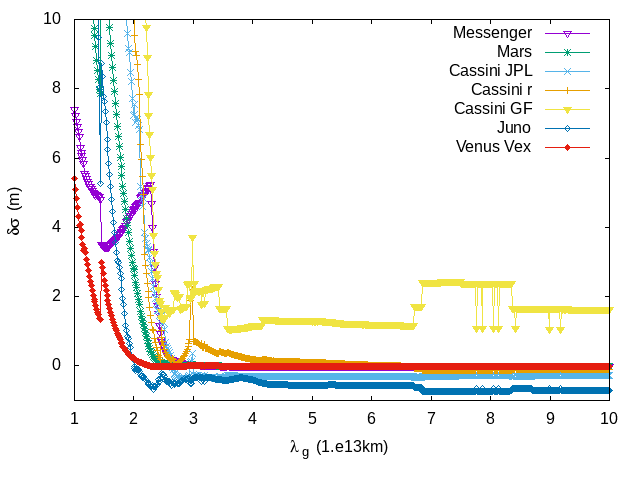}
	\caption{The standard dispersion of each sensitive data sets residuals have been plotted with respect to the Compton wavelength of the graviton. We have removed the standard dispersion of the reference solution residuals to each data sets (see Table \ref{tab:sensdata}). "Cassini JPL" corresponds to data provided by JPL, "Cassini r" corresponds to the Cassini Navigation and Gravity flybys data reduced by our team, "Cassini GF" corresponds to the Grand Finale of Cassini, and Mars correspond to a daily average of the data of Mars Odyssey and Mars Express.}
	\label{fig:residuals}
\end{figure}

We show the residual standard dispersions of the most sensitive observations in Fig. \ref{fig:residuals}. We note that, contrary to the results obtained with INPOP17a \cite{bernus:2019prl}, the considered residuals now degrade roughly simultaneously. However, as $\lg$ decreases, Cassini and Messenger are the first degraded residuals, the ones for Mars come just after, and Juno residuals are the least to be degraded. We note that the residuals of Cassini reduced by our team \cite{fienga:2019inpop,DiRuscio2020} increase around $\lg=3\times 10^{13}$km, then decrease, then , at last, increase significantly around $2.5\times 10^{13}$km. We observe the same behaviour, whereas less significantly, with the JPL Cassini data. Mars data has a more monotonic behavior.

Here we can see that the Pearson test that we used for INPOP17b in \cite{bernus:2019prl} would not have been possible with INPOP19a since several observational data sets have their residuals degraded at the same time. Indeed, the degradation of the residuals due to the Yukawa suppression is better distributed on the several data sets than in INPOP17b, thus a global analysis and a global criterion was necessary.

In Fig. \ref{fig:residuals}, we also see the evolution of the residual standard dispersion of Venus Vex mission. One can see that when they are significantly degraded for low values of $\lg$, all the other residuals are already high. One can deduce that Venus Vex mission will not contribute directly to constrain the Yukawa suppression. This example illustrates how we have chosen the sensitive observations and the non sensitive observations for INPOP19a. We have taken the example of Venus because it was the most sensitive set of data that was removed, but according to this process, more than half of the observations were not sensitive and removed. It is important to note that even if only the sensitive data were used for the statistical criterion, all of them were used for the adjustment procedure.

\begin{figure}
        \includegraphics[scale=0.6]{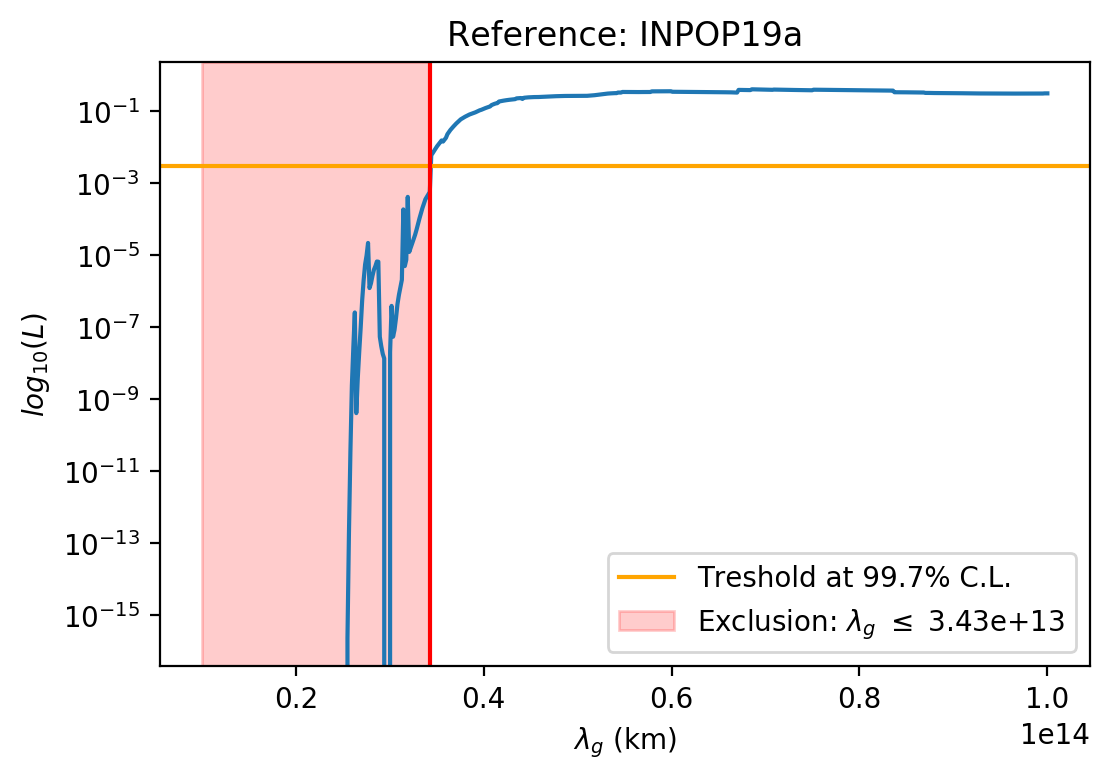}
        \caption{Likelihood for each value of $\lg$ of the Yukawa suppression from INPOP19a. 
The horizontal line represents the 99.7\% confidence limit; while the vertical line indicates the value of $\lg$ at that limit. The shaded area is the exclusion region. The constraint reads $\lg >$\Ltrois \ at the \cltrois. }
        \label{fig:likelihood}
\end{figure}

We then compute $L(z(\lg))$ for the selected data sets and we get Fig. \ref{fig:likelihood}. As expected, the results of the simulation tell us that $L(z(\lg))$ reaches very low values for low values of $\lg$ and converges to $1/2$ when $\lg$ tends to infinity.
 In order to be conservative, we first find the lowest value of $\lg$ such that $L(\lg)$ is still above a given threshold (e.g. 0.1 or 0.003). We then identify the next (lower) value to this one as our constraint. The results read $\lg >$ \Lnonente \ at the \clnonente, $\lg >$ \Ltrois \ at the \cltrois, and $\lg>$\Lcinq \ at the 99.99997$\%$ C.L. 
This is a significant improvement of our previous constraint. In particular, let us note that our \clnonente. constraint is now stronger than the one reported by the LIGO-Virgo collaboration in the radiative limit from the first gravitational-wave transient catalog \cite{ligo:2019prd}---which reads $\lg \geq 2.6 \times 10^{13}$km ($m_g \leq 4.7 \times 10^{-23}$ eV/$c^2$) at the \clnonente. This can be explained by the very good update of our planetary ephemeris INPOP from INPOP17b to INPOP19a---as it has been discussion in the introduction. As we have said in our previous work \cite{bernus:2019prl}, we remind that the fact that our constraint has the same order of magnitude that the one of LIGO-Virgo collaboration is a pure coincidence, because they check different phenomenology from different types of data: orbital in our study and radiative in LIGO-Virgo collaboration's work \cite{ligo:2019prd}.
Finally, one can notice that the fit procedure works better here than in INPOP17b, because most of the sensitive  data sets (Messenger, Mars range missions, Cassini, Juno) are degraded simultaneously for higher values of $\lg$, whereas in INPOP17b, only Cassini were significantly degraded \cite{bernus:2019prl}. This probably shows that the errors are more balanced between the different observations in INPOP19a.

\begin{figure}
        \includegraphics[scale=0.6]{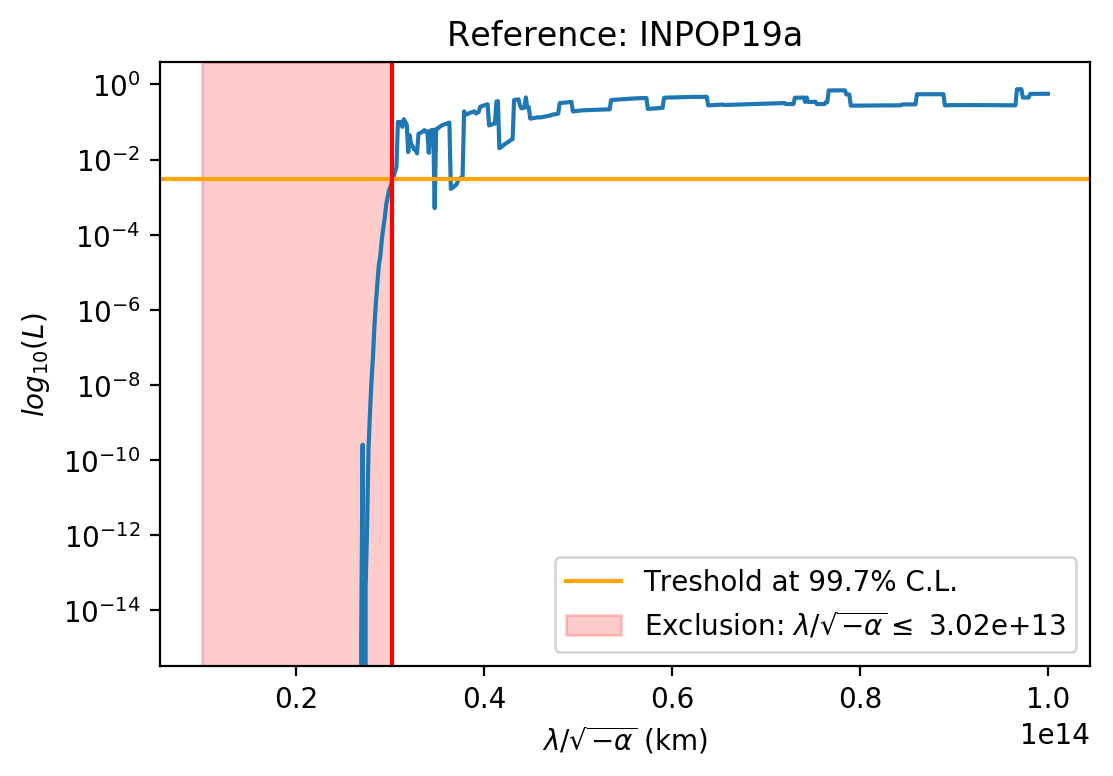}
        \caption{Likelihood with respect to the fifth force parameters $\lambda/\sqrt{-\alpha}$ for INPOP19a with negative values of $\alpha$. The horizontal line represents the 99.7\% confidence limit; while the vertical segment indicates the value of $\lg$ at that limit. The shaded area is the exclusion region. The constraint reads $\lambda / \sqrt{-\alpha} >$\LtroisNeg \ at the \cltrois.}
        \label{fig:gravmass_moins}
\end{figure}

In \cite{bernus:2019prl}, we noted that our result also constrains the \textit{fifth force formalism}. This force transforms the Newtonian potential into $V(r)=V_{Newton}(r)(1+\alpha\mathrm{exp}(-r/\lambda))$. Indeed, we had shown that our constraint on $\lg$ could be converted on a constraint on a combination of the fifth force parameters $\lambda/\sqrt{\alpha}$ if $\lg\gg r$ and $\alpha\ll 1$.  In order to encompass both negative and positive values of $\alpha$, we also performe the same numerical simulations with an opposite term for the additional acceleration of Yukawa suppression. We report the result in Fig. \ref{fig:gravmass_moins}. If $\alpha<0$, the limit reads $\lambda/\sqrt{|\alpha|} >$ \LnonenteNeg \ at the \clnonente, and $\lambda/\sqrt{|\alpha|} >$ \LtroisNeg \ at the \cltrois. Whereas if $\alpha > 0$, the limit reads $\lambda/\sqrt{|\alpha|} >$ \Lnonente \ at the \clnonente, $\lambda/\sqrt{|\alpha|} >$ \Ltrois \ at the \cltrois and $\lambda/\sqrt{|\alpha|}>$\LcinqNeg \ at the 99.99997$\%$ C.L. As one could have expected, one gets similar results for negative and positive values of $\alpha$. 

\begin{table*}[]
    \centering
        \caption{Constraints  at 90$\%$, 99.7 $\%$ and 99.99997\% C.L. for the Compton length $\lg$ in the case of the Yukawa suppression and for the ratio $\lambda/\sqrt{|\alpha|}$ for the Fifth force. For $\lg$, the constraints given in Columns 2 to 5 are minimum values when for $m_g$, the indicated constraints are maximum values. Values obtained with INPOP19a are given in Columns 4 to 6 when values obtained with INPOP17b \cite{bernus:2019prl} and by GWTC-1 from LIGO-Virgo collaboration \cite{ligo:2019prd} are  given for comparisons in Columns 2 and 3.}
    \begin{tabular}{c | l | l | l l l}
                                               &\multicolumn{1}{c |}{GWTC-1}& \multicolumn{1}{c |}{INPOP17a} & \multicolumn{2}{c}{INPOP19a}\\

                                               & 90$\%$ C.L.       & 90$\%$ C.L.                    & 90$\%$ C.L.    & 99.7 $\%$ C.L.  & 99.99997$\%$ C.L. \\
    \hline
       Yukawa suppression                      &                   &                                &                & \\
         $\lg$                                 &$2.6\times10^{13}$ km& \Lprl                        &\Lnonente       &  \Ltrois  & \Lcinq \\
         $m_g$                         &$4.7\times10^{-23}$eV$/c^2$& \mprl                          & \mnonente      &  \mtrois  &  \mcinq \\
    \hline
         Fifth Force                           &                   &                                &                &           &\\
    $\lambda/\sqrt{|\alpha|}$ for $\alpha > 0$ &                   & \Lprl                          &  \Lnonente     &  \Ltrois  &  \Lcinq\\
    $\lambda/\sqrt{|\alpha|}$ for  $\alpha < 0$&                   &                                &  \LnonenteNeg  &\LtroisNeg & \LcinqNeg\\
    \hline
    \end{tabular}
    \label{tab:resultats}
\end{table*}


\section{Conclusion}

We used a new statistical criterion to infer the constraint on a Yukawa suppression of the Newtonian potential---generically associated to a massive gravity theory---at the Solar system scales by using a global fit with planetary ephemeris. The new method is based on the global observational $\chi^2$ analysis and a likelihood criterion. After verifying that the new criterion led to the same result as our former (less general) one when applied to INPOP17b, we applied this analysis on our new ephemeris INPOP19a. The constraint reads $\lg>$ \Ltrois \ (or $m_g<$ \mtrois) with a 99.7\% confidence level. For the fifth force theory, if $\lambda\gg r$ and $\alpha\ll 1$ the constraint reads $\lambda/\sqrt{|\alpha|} > $ \LtroisNeg \ if $\alpha<0$, and $\lambda/\sqrt{|\alpha|} >$ \Ltrois \ if $\alpha>0$, with a 99.7\% confidence level.

This is a significant improvement with respect to our previous constraint with INPOP17b. In particular, we noted that our \clnonente. constraint is now stronger than the one given by the LIGO-Virgo collaboration in the radiative limit from the first gravitational-wave transient catalog \cite{ligo:2019prd}---although we wish to remind the reader that constraints from gravitational waves and from solar system ephemeris are complementary since they test different effects that could
be explained by a massive graviton. 
We expect further improvements in the future thanks to new data from spatial probes---in particular from Juno and BepiColombo in a near future---and to potential improvements of the the Solar System model. 



\begin{thebibliography}{16}%
\makeatletter
\providecommand \@ifxundefined [1]{%
 \@ifx{#1\undefined}
}%
\providecommand \@ifnum [1]{%
 \ifnum #1\expandafter \@firstoftwo
 \else \expandafter \@secondoftwo
 \fi
}%
\providecommand \@ifx [1]{%
 \ifx #1\expandafter \@firstoftwo
 \else \expandafter \@secondoftwo
 \fi
}%
\providecommand \natexlab [1]{#1}%
\providecommand \enquote  [1]{``#1''}%
\providecommand \bibnamefont  [1]{#1}%
\providecommand \bibfnamefont [1]{#1}%
\providecommand \citenamefont [1]{#1}%
\providecommand \href@noop [0]{\@secondoftwo}%
\providecommand \href [0]{\begingroup \@sanitize@url \@href}%
\providecommand \@href[1]{\@@startlink{#1}\@@href}%
\providecommand \@@href[1]{\endgroup#1\@@endlink}%
\providecommand \@sanitize@url [0]{\catcode `\\12\catcode `\$12\catcode
  `\&12\catcode `\#12\catcode `\^12\catcode `\_12\catcode `\%12\relax}%
\providecommand \@@startlink[1]{}%
\providecommand \@@endlink[0]{}%
\providecommand \url  [0]{\begingroup\@sanitize@url \@url }%
\providecommand \@url [1]{\endgroup\@href {#1}{\urlprefix }}%
\providecommand \urlprefix  [0]{URL }%
\providecommand \Eprint [0]{\href }%
\providecommand \doibase [0]{http://dx.doi.org/}%
\providecommand \selectlanguage [0]{\@gobble}%
\providecommand \bibinfo  [0]{\@secondoftwo}%
\providecommand \bibfield  [0]{\@secondoftwo}%
\providecommand \translation [1]{[#1]}%
\providecommand \BibitemOpen [0]{}%
\providecommand \bibitemStop [0]{}%
\providecommand \bibitemNoStop [0]{.\EOS\space}%
\providecommand \EOS [0]{\spacefactor3000\relax}%
\providecommand \BibitemShut  [1]{\csname bibitem#1\endcsname}%
\let\auto@bib@innerbib\@empty
\bibitem [{\citenamefont {{Will}}(1998)}]{will:1998pr}%
  \BibitemOpen
  \bibfield  {author} {\bibinfo {author} {\bibfnamefont {C.~M.}\ \bibnamefont
  {{Will}}},\ }\bibfield  {title} {\enquote {\bibinfo {title} {{Bounding the
  mass of the graviton using gravitational-wave observations of inspiralling
  compact binaries}},}\ }\href {\doibase 10.1103/PhysRevD.57.2061} {\bibfield
  {journal} {\bibinfo  {journal} {\prd}\ }\textbf {\bibinfo {volume} {57}},\
  \bibinfo {pages} {2061--2068} (\bibinfo {year} {1998})},\ \Eprint
  {http://arxiv.org/abs/gr-qc/9709011} {gr-qc/9709011} \BibitemShut {NoStop}%
\bibitem [{\citenamefont {{de Rham}}(2014)}]{derham:2014lr}%
  \BibitemOpen
  \bibfield  {author} {\bibinfo {author} {\bibfnamefont {C.}~\bibnamefont {{de
  Rham}}},\ }\bibfield  {title} {\enquote {\bibinfo {title} {{Massive
  Gravity}},}\ }\href {\doibase 10.12942/lrr-2014-7} {\bibfield  {journal}
  {\bibinfo  {journal} {Living Reviews in Relativity}\ }\textbf {\bibinfo
  {volume} {17}},\ \bibinfo {eid} {7} (\bibinfo {year} {2014})},\ \Eprint
  {http://arxiv.org/abs/1401.4173} {arXiv:1401.4173 [hep-th]} \BibitemShut
  {NoStop}%
\bibitem [{\citenamefont {{de Rham}}\ \emph {et~al.}(2017)\citenamefont {{de
  Rham}}, \citenamefont {{Deskins}}, \citenamefont {{Tolley}},\ and\
  \citenamefont {{Zhou}}}]{derham:2017rp}%
  \BibitemOpen
  \bibfield  {author} {\bibinfo {author} {\bibfnamefont {C.}~\bibnamefont {{de
  Rham}}}, \bibinfo {author} {\bibfnamefont {J.~T.}\ \bibnamefont {{Deskins}}},
  \bibinfo {author} {\bibfnamefont {A.~J.}\ \bibnamefont {{Tolley}}}, \ and\
  \bibinfo {author} {\bibfnamefont {S.-Y.}\ \bibnamefont {{Zhou}}},\ }\bibfield
   {title} {\enquote {\bibinfo {title} {{Graviton mass bounds}},}\ }\href
  {\doibase 10.1103/RevModPhys.89.025004} {\bibfield  {journal} {\bibinfo
  {journal} {Reviews of Modern Physics}\ }\textbf {\bibinfo {volume} {89}},\
  \bibinfo {eid} {025004} (\bibinfo {year} {2017})},\ \Eprint
  {http://arxiv.org/abs/1606.08462} {arXiv:1606.08462} \BibitemShut {NoStop}%
\bibitem [{\citenamefont {Bernus}\ \emph {et~al.}(2019)\citenamefont {Bernus},
  \citenamefont {Minazzoli}, \citenamefont {Fienga}, \citenamefont {Gastineau},
  \citenamefont {Laskar},\ and\ \citenamefont {Deram}}]{bernus:2019prl}%
  \BibitemOpen
  \bibfield  {author} {\bibinfo {author} {\bibfnamefont {L.}~\bibnamefont
  {Bernus}}, \bibinfo {author} {\bibfnamefont {O.}~\bibnamefont {Minazzoli}},
  \bibinfo {author} {\bibfnamefont {A.}~\bibnamefont {Fienga}}, \bibinfo
  {author} {\bibfnamefont {M.}~\bibnamefont {Gastineau}}, \bibinfo {author}
  {\bibfnamefont {J.}~\bibnamefont {Laskar}}, \ and\ \bibinfo {author}
  {\bibfnamefont {P.}~\bibnamefont {Deram}},\ }\bibfield  {title} {\enquote
  {\bibinfo {title} {Constraining the mass of the graviton with the planetary
  ephemeris inpop},}\ }\href {\doibase 10.1103/PhysRevLett.123.161103}
  {\bibfield  {journal} {\bibinfo  {journal} {Phys. Rev. Lett.}\ }\textbf
  {\bibinfo {volume} {123}},\ \bibinfo {pages} {161103} (\bibinfo {year}
  {2019})}\BibitemShut {NoStop}%
\bibitem [{\citenamefont {{Fienga}}\ \emph {et~al.}(2008)\citenamefont
  {{Fienga}}, \citenamefont {{Manche}}, \citenamefont {{Laskar}},\ and\
  \citenamefont {{Gastineau}}}]{fienga:2008aa}%
  \BibitemOpen
  \bibfield  {author} {\bibinfo {author} {\bibfnamefont {A.}~\bibnamefont
  {{Fienga}}}, \bibinfo {author} {\bibfnamefont {H.}~\bibnamefont {{Manche}}},
  \bibinfo {author} {\bibfnamefont {J.}~\bibnamefont {{Laskar}}}, \ and\
  \bibinfo {author} {\bibfnamefont {M.}~\bibnamefont {{Gastineau}}},\
  }\bibfield  {title} {\enquote {\bibinfo {title} {{INPOP06: a new numerical
  planetary ephemeris}},}\ }\href {\doibase 10.1051/0004-6361:20066607}
  {\bibfield  {journal} {\bibinfo  {journal} {\aap}\ }\textbf {\bibinfo
  {volume} {477}},\ \bibinfo {pages} {315--327} (\bibinfo {year}
  {2008})}\BibitemShut {NoStop}%
\bibitem [{\citenamefont {Moyer}(2003)}]{moyer:2003}%
  \BibitemOpen
  \bibfield  {author} {\bibinfo {author} {\bibfnamefont {T.~D.}\ \bibnamefont
  {Moyer}},\ }\href {\doibase 10.1002/0471728470} {\emph {\bibinfo {title}
  {Deep Space Communications and Navigation Series}}},\ Vol.~\bibinfo {volume}
  {2}\ (\bibinfo  {publisher} {John Wiley {\&} Sons, Inc.},\ \bibinfo {address}
  {Hoboken, NJ, USA},\ \bibinfo {year} {2003})\BibitemShut {NoStop}%
\bibitem [{\citenamefont {{Fienga}}\ \emph {et~al.}(2011)\citenamefont
  {{Fienga}}, \citenamefont {{Laskar}}, \citenamefont {{Kuchynka}},
  \citenamefont {{Manche}}, \citenamefont {{Desvignes}}, \citenamefont
  {{Gastineau}}, \citenamefont {{Cognard}},\ and\ \citenamefont
  {{Theureau}}}]{fienga:2011cm}%
  \BibitemOpen
  \bibfield  {author} {\bibinfo {author} {\bibfnamefont {A.}~\bibnamefont
  {{Fienga}}}, \bibinfo {author} {\bibfnamefont {J.}~\bibnamefont {{Laskar}}},
  \bibinfo {author} {\bibfnamefont {P.}~\bibnamefont {{Kuchynka}}}, \bibinfo
  {author} {\bibfnamefont {H.}~\bibnamefont {{Manche}}}, \bibinfo {author}
  {\bibfnamefont {G.}~\bibnamefont {{Desvignes}}}, \bibinfo {author}
  {\bibfnamefont {M.}~\bibnamefont {{Gastineau}}}, \bibinfo {author}
  {\bibfnamefont {I.}~\bibnamefont {{Cognard}}}, \ and\ \bibinfo {author}
  {\bibfnamefont {G.}~\bibnamefont {{Theureau}}},\ }\bibfield  {title}
  {\enquote {\bibinfo {title} {{The INPOP10a planetary ephemeris and its
  applications in fundamental physics}},}\ }\href {\doibase
  10.1007/s10569-011-9377-8} {\bibfield  {journal} {\bibinfo  {journal}
  {Celestial Mechanics and Dynamical Astronomy}\ }\textbf {\bibinfo {volume}
  {111}},\ \bibinfo {pages} {363--385} (\bibinfo {year} {2011})},\ \Eprint
  {http://arxiv.org/abs/1108.5546} {arXiv:1108.5546 [astro-ph.EP]} \BibitemShut
  {NoStop}%
\bibitem [{\citenamefont {{Fienga}}\ \emph {et~al.}(2015)\citenamefont
  {{Fienga}}, \citenamefont {{Laskar}}, \citenamefont {{Exertier}},
  \citenamefont {{Manche}},\ and\ \citenamefont {{Gastineau}}}]{fienga:2015cm}%
  \BibitemOpen
  \bibfield  {author} {\bibinfo {author} {\bibfnamefont {A.}~\bibnamefont
  {{Fienga}}}, \bibinfo {author} {\bibfnamefont {J.}~\bibnamefont {{Laskar}}},
  \bibinfo {author} {\bibfnamefont {P.}~\bibnamefont {{Exertier}}}, \bibinfo
  {author} {\bibfnamefont {H.}~\bibnamefont {{Manche}}}, \ and\ \bibinfo
  {author} {\bibfnamefont {M.}~\bibnamefont {{Gastineau}}},\ }\bibfield
  {title} {\enquote {\bibinfo {title} {{Numerical estimation of the sensitivity
  of INPOP planetary ephemerides to general relativity parameters}},}\ }\href
  {\doibase 10.1007/s10569-015-9639-y} {\bibfield  {journal} {\bibinfo
  {journal} {Celestial Mechanics and Dynamical Astronomy}\ }\textbf {\bibinfo
  {volume} {123}},\ \bibinfo {pages} {325--349} (\bibinfo {year}
  {2015})}\BibitemShut {NoStop}%
\bibitem [{\citenamefont {{Fienga}}\ \emph {et~al.}(2019)\citenamefont
  {{Fienga}}, \citenamefont {{Deram}}, \citenamefont {{Viswanathan}},
  \citenamefont {{Di Ruscio}}, \citenamefont {{Bernus}}, \citenamefont
  {{Durante}}, \citenamefont {{Gastineau}},\ and\ \citenamefont
  {{Laskar}}}]{fienga:2019inpop}%
  \BibitemOpen
  \bibfield  {author} {\bibinfo {author} {\bibfnamefont {A.}~\bibnamefont
  {{Fienga}}}, \bibinfo {author} {\bibfnamefont {P.}~\bibnamefont {{Deram}}},
  \bibinfo {author} {\bibfnamefont {V.}~\bibnamefont {{Viswanathan}}}, \bibinfo
  {author} {\bibfnamefont {A.}~\bibnamefont {{Di Ruscio}}}, \bibinfo {author}
  {\bibfnamefont {L.}~\bibnamefont {{Bernus}}}, \bibinfo {author}
  {\bibfnamefont {D.}~\bibnamefont {{Durante}}}, \bibinfo {author}
  {\bibfnamefont {M.}~\bibnamefont {{Gastineau}}}, \ and\ \bibinfo {author}
  {\bibfnamefont {J.}~\bibnamefont {{Laskar}}},\ }\bibfield  {title} {\enquote
  {\bibinfo {title} {{INPOP19a planetary ephemeris}},}\ }\href@noop {}
  {\bibfield  {journal} {\bibinfo  {journal} {Notes Scientifiques et Techniques
  de l'Institut de Mecanique Celeste}\ }\textbf {\bibinfo {volume} {109}}
  (\bibinfo {year} {2019})}\BibitemShut {NoStop}%
\bibitem [{\citenamefont {{Hees}}\ \emph {et~al.}(2014)\citenamefont {{Hees}},
  \citenamefont {{Folkner}}, \citenamefont {{Jacobson}},\ and\ \citenamefont
  {{Park}}}]{hees2014prd}%
  \BibitemOpen
  \bibfield  {author} {\bibinfo {author} {\bibfnamefont {A.}~\bibnamefont
  {{Hees}}}, \bibinfo {author} {\bibfnamefont {W.~M.}\ \bibnamefont
  {{Folkner}}}, \bibinfo {author} {\bibfnamefont {R.~A.}\ \bibnamefont
  {{Jacobson}}}, \ and\ \bibinfo {author} {\bibfnamefont {R.~S.}\ \bibnamefont
  {{Park}}},\ }\bibfield  {title} {\enquote {\bibinfo {title} {{Constraints on
  modified Newtonian dynamics theories from radio tracking data of the Cassini
  spacecraft}},}\ }\href {\doibase 10.1103/PhysRevD.89.102002} {\bibfield
  {journal} {\bibinfo  {journal} {\prd}\ }\textbf {\bibinfo {volume} {89}},\
  \bibinfo {eid} {102002} (\bibinfo {year} {2014})},\ \Eprint
  {http://arxiv.org/abs/1402.6950} {arXiv:1402.6950 [gr-qc]} \BibitemShut
  {NoStop}%
\bibitem [{\citenamefont {{Di Ruscio}}\ \emph {et~al.}(2020)\citenamefont {{Di
  Ruscio}}, \citenamefont {{Fienga}}, \citenamefont {{Durante}}, \citenamefont
  {{Iess}},\ and\ \citenamefont {{Laskar}}}]{DiRuscio2020}%
  \BibitemOpen
  \bibfield  {author} {\bibinfo {author} {\bibfnamefont {A.}~\bibnamefont {{Di
  Ruscio}}}, \bibinfo {author} {\bibfnamefont {A.}~\bibnamefont {{Fienga}}},
  \bibinfo {author} {\bibfnamefont {D.}~\bibnamefont {{Durante}}}, \bibinfo
  {author} {\bibfnamefont {L.}~\bibnamefont {{Iess}}}, \ and\ \bibinfo {author}
  {\bibfnamefont {M.}~\bibnamefont {{Laskar}}, \bibfnamefont
  {J.and~{Gastineau}}},\ }\bibfield  {title} {\enquote {\bibinfo {title} {{An
  estimate of the Kuiper belt mass from Cassini tracking data and INPOP19a}},}\
  }\href@noop {} {\bibfield  {journal} {\bibinfo  {journal} {\aap (in press)}\
  } (\bibinfo {year} {2020})}\BibitemShut {NoStop}%
\bibitem [{\citenamefont {{Fienga}}\ \emph {et~al.}(2020)\citenamefont
  {{Fienga}}, \citenamefont {{Di Ruscio}}, \citenamefont {{Bernus}},
  \citenamefont {{Deram}}, \citenamefont {{Durante}},\ and\ \citenamefont
  {{Laskar}}}]{fienga:2020aap}%
  \BibitemOpen
  \bibfield  {author} {\bibinfo {author} {\bibfnamefont {A.}~\bibnamefont
  {{Fienga}}}, \bibinfo {author} {\bibfnamefont {A.}~\bibnamefont {{Di
  Ruscio}}}, \bibinfo {author} {\bibfnamefont {L.}~\bibnamefont {{Bernus}}},
  \bibinfo {author} {\bibfnamefont {P.}~\bibnamefont {{Deram}}}, \bibinfo
  {author} {\bibfnamefont {D.}~\bibnamefont {{Durante}}}, \ and\ \bibinfo
  {author} {\bibfnamefont {L.}~\bibnamefont {{Laskar}}, \bibfnamefont
  {J.and~{Iess}}},\ }\bibfield  {title} {\enquote {\bibinfo {title} {{New
  constraints on P9 localisation obtained with the INPOP19aplanetary
  ephemeris}},}\ }\href@noop {} {\bibfield  {journal} {\bibinfo  {journal}
  {\aap (in press)}\ } (\bibinfo {year} {2020})}\BibitemShut {NoStop}%
\bibitem [{\citenamefont {Will}(2018)}]{will:2018cq}%
  \BibitemOpen
  \bibfield  {author} {\bibinfo {author} {\bibfnamefont {Clifford~M}\
  \bibnamefont {Will}},\ }\bibfield  {title} {\enquote {\bibinfo {title} {Solar
  system versus gravitational-wave bounds on the graviton mass},}\ }\href
  {http://stacks.iop.org/0264-9381/35/i=17/a=17LT01} {\bibfield  {journal}
  {\bibinfo  {journal} {Classical and Quantum Gravity}\ }\textbf {\bibinfo
  {volume} {35}},\ \bibinfo {pages} {17LT01} (\bibinfo {year}
  {2018})}\BibitemShut {NoStop}%
\bibitem [{\citenamefont {{Viswanathan}}\ \emph {et~al.}(2017)\citenamefont
  {{Viswanathan}}, \citenamefont {{Fienga}}, \citenamefont {{Gastineau}},\ and\
  \citenamefont {{Laskar}}}]{viswanathan:2018dc}%
  \BibitemOpen
  \bibfield  {author} {\bibinfo {author} {\bibfnamefont {V.}~\bibnamefont
  {{Viswanathan}}}, \bibinfo {author} {\bibfnamefont {A.}~\bibnamefont
  {{Fienga}}}, \bibinfo {author} {\bibfnamefont {M.}~\bibnamefont
  {{Gastineau}}}, \ and\ \bibinfo {author} {\bibfnamefont {J.}~\bibnamefont
  {{Laskar}}},\ }\bibfield  {title} {\enquote {\bibinfo {title} {{INPOP17a
  planetary ephemerides}},}\ }\href@noop {} {\bibfield  {journal} {\bibinfo
  {journal} {Notes Scientifiques et Techniques de l'Institut de Mecanique
  Celeste}\ }\textbf {\bibinfo {volume} {108}} (\bibinfo {year} {2017})},\
  \bibinfo {note} {last Accessed: 2018-11-13}\BibitemShut {NoStop}%
\bibitem [{\citenamefont {{Verma}}\ and\ \citenamefont
  {{Margot}}(2016)}]{verma2016jgr}%
  \BibitemOpen
  \bibfield  {author} {\bibinfo {author} {\bibfnamefont {A.~K.}\ \bibnamefont
  {{Verma}}}\ and\ \bibinfo {author} {\bibfnamefont {J.-L.}\ \bibnamefont
  {{Margot}}},\ }\bibfield  {title} {\enquote {\bibinfo {title} {{Mercury's
  gravity, tides, and spin from MESSENGER radio science data}},}\ }\href
  {\doibase 10.1002/2016JE005037} {\bibfield  {journal} {\bibinfo  {journal}
  {Journal of Geophysical Research (Planets)}\ }\textbf {\bibinfo {volume}
  {121}},\ \bibinfo {pages} {1627--1640} (\bibinfo {year} {2016})},\ \Eprint
  {http://arxiv.org/abs/1608.01360} {arXiv:1608.01360 [astro-ph.EP]}
  \BibitemShut {NoStop}%
\bibitem [{\citenamefont {{The LIGO Scientific Collaboration}}\ and\
  \citenamefont {{the Virgo Collaboration}}(2019)}]{ligo:2019prd}%
  \BibitemOpen
  \bibfield  {author} {\bibinfo {author} {\bibnamefont {{The LIGO Scientific
  Collaboration}}}\ and\ \bibinfo {author} {\bibnamefont {{the Virgo
  Collaboration}}} (\bibinfo {collaboration} {The LIGO Scientific Collaboration
  and the Virgo Collaboration}),\ }\bibfield  {title} {\enquote {\bibinfo
  {title} {Tests of general relativity with the binary black hole signals from
  the ligo-virgo catalog gwtc-1},}\ }\href {\doibase
  10.1103/PhysRevD.100.104036} {\bibfield  {journal} {\bibinfo  {journal}
  {Phys. Rev. D}\ }\textbf {\bibinfo {volume} {100}},\ \bibinfo {pages}
  {104036} (\bibinfo {year} {2019})}\BibitemShut {NoStop}%
\end{thebibliography}


%
\end{document}